\documentstyle[preprint,aps]{revtex}
\begin{document}
\draft
\title{Hadronic cross-section fluctuations and proton coherent
diffractive dissociation on helium}
\author{M.Strikman}
\address{Department of Physics\\
The Pennsylvania State University, University Park, PA 16802\\
and Inst. of Nuclear Physics, St. Petersburg, Russia}
\author{and\\
V.Guzey}
\address{Department of Physics\\
The Pennsylvania State University, University Park, PA 16802}
\maketitle
\begin{abstract}

 The differential cross-section  of inelastic coherent diffractive
dissociation off nuclei $p~+~^4$He $\rightarrow X ~+ ~^4$He
is expressed in terms of the relative cumulants
of the cross-section distribution $P_N(\sigma)$.
The theoretical result for the ratio
$r=\Bigg(\frac{d\sigma_{diff}}{dt}\Bigg)^{pHe}_{t=0} \Bigg/
 \Bigg(\frac{d\sigma_{diff}}{dt}\Bigg)^{pp}_{t=0}= 6.8 \div 7.6$ is close to
the value $r=7.1 \pm 0.7$ which we extracted from the FNAL data.
These are the only $A>2$ data of this kind. The comparison provides the
first confirmation of the cross-section fluctuation approach to the
description of the absolute
value of the inelastic diffraction cross-section off nuclei. It
 provides also a new constraint on the first 4  cumulants of the
cross-section distribution.
\end{abstract}

\newpage

\narrowtext

\section{Introduction}
One of manifestations of the composite structure of hadrons is that
constituents of rather  different size are present in hadrons.
At high energies transition time from one configuration to another
exceeds by far the time of the interaction with the target:
\begin{equation}
\frac{2p_{lab}}{M^2-m^2} \gg 2R.
\end{equation}
Here $M$ is the mass of an inelastic cross-section eigenstate and $m$ is the
ground- state mass of the hadron; $R$ is a longitudinal length characterizing
the interaction of the target.
Hence one can consider these fluctuations as frozen during the collision
time and then integrate over probability of configurations in a
projectile. Since these configurations
interact with very different strength one should take the fluctuation of
 the interaction strength  -- cross-section fluctuations  -- into account
 in a realistic picture of the hadron - nucleus interactions, for a recent
 review
see \cite{FMS}.

The convenient formalism to accommodate this physics -- the
 scattering eigenstate formalism was suggested long ago by
 Good and Walker~\cite{GW}. The projectile wave function $|\Psi\rangle$
is to be expanded as a sum of eigenstates of the purely imaginary T-matrix
(for simplicity
we consider scattering amplitude as purely imaginary)
\begin{equation}
|\Psi\rangle=\sum_{\kappa}c_{\kappa}|\psi_{\kappa}\rangle, \qquad
 \sum_{\kappa}|c_{\kappa}|^2=1,
\end{equation}
provided
\begin{equation}
Im\,T|\psi_{\kappa}\rangle=T_{\kappa}|\psi_{\kappa}\rangle.
\end{equation}
$T_{\kappa}$ is the cross-section for the interaction of projectile
configuration $k$ with the target.
In this basis there are no transitions between different states and
 this makes it possible to describe a number of  physical processes
 in terms of a distribution over the values of the cross-section, $P(\sigma)$.

Namely, $P(\sigma)=\sum_{\kappa}|c_{\kappa}|^2\delta(\sigma-T_{\kappa})$
 gives the probability that a given configuration interacts with a nucleon
 with a total cross-section $\sigma$. It allows us to treat the projectile
 as a coherent superposition of scattering eigenstates, each with an
 eigenvalue $\sigma$. This idea gives rise to the term {\em hadronic
 cross-section fluctuations}.

In refs.\cite{BBFS,BBFS1}  $P(\sigma)$ was determined for pion and nucleon
projectiles using data on diffraction off nucleons and deuterons as well
as the analog of the quark counting rules to fix the  behavior of  $P(\sigma)$
for small  $\sigma$. Besides,  $P_{\pi}(\sigma \ll \left< \sigma \right>)$
was calculated directly in QCD \cite{BBFS}.
More recently a similar technique was used to calculate the process
of electroproduction of $\rho$-mesons by longitudinally polarized virtual
photons \cite{Brodskyetal}. The predictions of Ref.\cite{Brodskyetal}
 were very recently confirmed by the ZEUS experiment at HERA \cite{ZEUS},
thus providing indirect
confirmation of the calculation of
 $P_{\pi}(\sigma \ll \left< \sigma \right>)$.

It was suggested in Ref.\cite{FMS93} that the data on inelastic
 coherent diffraction off nuclei would
provide a critical test of the concept of cross section fluctuations.
The total cross sections of diffraction dissociation were calculated.
However practically no data on the total cross section of coherent
 diffractive dissociation are available and only comparison of the
 predicted $A$-dependence of this cross section with the $A-$dependence of
exclusive channels measured at FNAL was possible.

However there exist previously overlooked unique data on the
 the process of proton inelastic diffractive scattering off $^4$He which
 were obtained nearly 15 years ago in the FNAL jet target experiment
 \cite{Bujak}.
So in this paper we will analyze these data to obtain another test of the
discussed approach. The specific feature of $^4$He nucleus
 is that its radius is small so as distinct from the approach used in
 Ref.\cite{FMS93}
 we cannot neglect the slope of diffractive amplitude, $\beta$,
 as compared to the slope of the $^4$He form-factors.

Diffractive scattering occurs when the final state has the same quantum
 numbers as the incident hadron $h$; that is, whenever it overlap
s any $|\psi_{\kappa}\rangle$. Thus, subtracting the elastic contribution,
 we can write
\begin{equation}
\Bigg(\frac{d\sigma_{diff}}{dt}\Bigg)^{hHe}=\frac{1}{16\pi} \sum_{\kappa}
|c_{\kappa}|^2T_{\kappa}^2-\Big(\sum_{\kappa}|c_{\kappa}|^2T_{\kappa}\Big)^2.
\end{equation}
Here\ $T_{\kappa}$ is the elastic scattering amplitude for a hadron
cross-section
eigenstate $|\psi_{\kappa}\rangle$ scattering off a nucleus of $^4$He.
 This formula enables us to investigate the relation between the differential
 cross-section, which can be extracted from experimental data, and the moments
 of the distribution $P(\sigma)$ which describe the cross-section fluctuations
 and contain information  about the hadronic structure.


\section{Differential cross-section}

When the instantaneous configuration can be considered frozen, the scattering
 process should be calculated first for the particular configuration and
 then integrated over all configurations which satisfy eq.(1), weighted
 by the probability of the configuration. In doing so one essentially
uses completeness of the intermediate and final states.

Let us suppose that the incident hadron in state $\kappa$ scatters off the
 nucleus of $^4$He in states $\imath$, $\jmath$, $\ell$, $m$ of its four
nucleons. Then,  to determine the scattering amplitude of this process,
 $F^{\kappa}_{\imath \jmath \ell m}$, we employed the Glauber method.
 This method requires knowledge of the amplitude of the hadron-hadron
scattering
 and the wave function of $^4$He. The parameterization of the hadron-hadron
 amplitude was taken as
\begin{equation}
f_{\imath}^{\kappa}(\vec{k}-\vec{k^{\prime}})=\frac{\imath}
{4\pi}\sigma_{\imath}^{\kappa}e^{-\frac{\beta}{2}(\vec{k}-\vec{k^{\prime}})^2}.
\end{equation}
Here $\sigma_{\imath}^{\kappa}$ is the total scattering cross-section for
 the hadron and nucleon in configurations $\kappa$ and $\imath$ respectively;
$\beta$ is a parameter whose numerical value will be discussed further below.
 The wave function of $^4$He, $\Phi$, was taken in a simple form \cite{LS}
\begin{equation}
\Phi=B\prod_{\imath=1}^{4}exp(-\alpha\cdot p_{\imath}^2)
\cdot\delta(\sum_{\imath=1}^{4}p_{\imath}),
\end{equation}
with $\alpha$=23(Gev/c)$^2$. It leads to the single nucleon form-factor
$F_{^4{\rm He}}(q^2)=exp(-3/8\alpha q^2)$.

 This form of the wave function allows us to reproduce well
the  total  cross section of $p^4$He scattering as well as the elastic cross
 section at small $t$.

For given instantaneous configurations of the projectile and
the target, and at zero transverse momentum, the Glauber method
leads to $F^{\kappa}_{\imath \jmath \ell m}$

\begin{equation}
Im\,F^{\kappa}_{\imath, \jmath, \ell, m}=\frac{\sigma^{\kappa}_{\imath}}
{\pi}-\frac{\sigma^{\kappa}_{\imath}\sigma^{\kappa}_{\jmath}}{16\pi^2
(\alpha+\beta)}+\frac{\sigma^{\kappa}_{\imath}\sigma^{\kappa}_{\jmath
}\sigma^{\kappa}_{\ell}}{48\pi^3(\alpha+\beta)^2}-\frac{\sigma^
{\kappa}_{\imath}\sigma^{\kappa}_{\jmath}\sigma^{\kappa}_{\ell}
\sigma^{\kappa}_{m}}{1024\pi^4(\alpha+\beta)^3}.
\end{equation}
One can see from eq.(7) that in the
$^4$He case one cannot neglect the slope of the rescattering amplitude
as compared to the slope of the nucleus many body form-factor, $\beta/
\alpha\approx$0.5 for the case of a nucleon projectile.
 At the same   time the parameter $\left<\beta\right> +\alpha$ is
 sufficiently large  as compared to the possible changes of $\beta$
 related to  the fluctuations of $\sigma$.
So we can neglect fluctuations of $\beta$.

After averaging over the configurations of the target,
\begin{equation}
Im\,T_{\kappa}=4\pi\sum_{\imath \jmath \ell m}|c_{\imath}|^2|c_{\jmath}|^2
|c_{\ell}|^2|c_{m}|^2 Im F^{\kappa}_{\imath \jmath \ell m},
\end{equation}
we obtain the elastic amplitude for a hadron cross-section eigenstate
$|\psi_{\kappa}\rangle$ scattering off $^4$He
\begin{equation}
Im \,T_{\kappa}=4\sigma_{\kappa}-\frac{3}{4\pi(\alpha+\beta)}
\sigma_{\kappa}^2+\frac{1}{12\pi^2(\alpha+\beta)^2}\sigma_{\kappa}^3-
\frac{1}{256\pi^3(\alpha+\beta)^3}\sigma_{\kappa}^4 .
\end{equation}
Thus we expressed the eigenvalues of the scattering eigenstate $|\psi_
{\kappa}\rangle$ for the interaction with $^4$He in terms of the eigenvalues
 $\sigma_{\kappa}$ of the interaction of $|\psi_{\kappa}$ with a nucleon.
 Therefore, from eq.(4) we see that the
differential cross-section is given by
\begin{eqnarray}
		    \Bigg(\frac{d\sigma_{diff}}{dt}\Bigg)^{hHe}_{t=0}&=&
{1 \over 16 \pi}
\Bigg\{16\bigg(\langle\sigma^2\rangle-\langle\sigma\rangle^2\bigg)
-\frac{6\langle\sigma\rangle^3}{\pi(\alpha+\beta)}\bigg
(\frac{\langle\sigma^3\rangle}{\langle\sigma\rangle^3}
-\frac{\langle\sigma^2\rangle}{\langle\sigma\rangle^2}\bigg) \nonumber\\
\nonumber\\
&+&\frac{59\langle\sigma\rangle^4}{48\pi^2(\alpha+\beta)^2}
\bigg(\frac{\langle\sigma^4\rangle}{\langle\sigma\rangle^4}
-\frac{27}{59}\frac{\langle\sigma^2\rangle^2}{\langle\sigma\rangle^4}
-\frac{32}{59}\frac{\langle\sigma^3\rangle}{\langle\sigma\rangle^3}\bigg)
 \nonumber\\
\nonumber\\
&-&\frac{5\langle\sigma\rangle^5}{32\pi^3(\alpha+\beta)^3}\bigg(\frac
{\langle\sigma^5\rangle}{\langle\sigma\rangle^5}-\frac{4}{5}\frac
{\langle\sigma^3\rangle\cdot\langle\sigma^2\rangle}{\langle\sigma\rangle^5}
-\frac{1}{5}\frac{\langle\sigma^4\rangle}{\langle\sigma\rangle^4}\bigg)\Bigg\}.
\end{eqnarray}
Here we neglected terms proportional to $\frac{\langle\sigma\rangle^6}
{(\alpha+\beta)^4},\cdots,\frac{\langle\sigma\rangle^8}{(\alpha+\beta)^6}$
 with an accuracy of 3\%.

Similarly, the differential cross-section of hadron-hadron scattering can
 be written in terms of averaging over internal configurations of the hadron
\cite{Pumplin}
\begin{equation}
\Bigg(\frac{d\sigma_{diff}}{dt}\Bigg)^{hh}_{t=0}=\frac{1}{16\pi}
\Bigg\{\langle\sigma^2\rangle-\langle\sigma\rangle^2\Bigg\}.
\end{equation}
 It is convenient to introduce the ratio of differential cross-sections $r$
\begin{equation}
r=\Bigg(\frac{d\sigma_{diff}}{dt}\Bigg)^{pHe}_{t=0} \Bigg/
 \Bigg(\frac{d\sigma_{diff}}{dt}\Bigg)^{pp}_{t=0}.
\end{equation}
Defining a factor  $\gamma$ and the second cumulant $\kappa_2$ as
\begin{equation}
\gamma=\frac{\langle\sigma\rangle}{\pi(\alpha+\beta)}, \qquad
\kappa_2=\frac{\langle\sigma^2\rangle-\langle\sigma\rangle^2}
{\langle\sigma\rangle^2},
\end{equation}
  we can write the ratio $r$ as
\begin{eqnarray}
r=16-\frac{6\gamma}{\kappa_2}\Bigg(\frac{\langle\sigma^3\rangle}
{\langle\sigma\rangle^3}-\frac{\langle\sigma^2\rangle}{\langle\sigma
\rangle^2}\Bigg)+\frac{59\gamma^2}{48\kappa_{2}}\Bigg(\frac{\langle
\sigma^4\rangle}{\langle\sigma\rangle^4}&-&\frac{27}{59}\frac{\langle
\sigma^2\rangle^2}{\langle\sigma\rangle^4}-\frac{32}{59}\frac{\langle
\sigma^3\rangle}{\langle\sigma\rangle^3}\Bigg)\nonumber\\
\nonumber\\       &-&\frac{5\gamma^3}{32\kappa_{2}}\Bigg(\frac{\langle
\sigma^5\rangle}{\langle\sigma\rangle^5}-\frac{4}{5}\frac{\langle\sigma^3
\rangle\langle\sigma^2\rangle}{\langle\sigma\rangle^5}-\frac{1}{5}
\frac{\langle\sigma^4\rangle}{\langle\sigma\rangle^4}\Bigg).
\end{eqnarray}

\section{ the cross-section distribution
 function and  experimental data}

 For numerical analysis we consider the case of a proton projectile for
which data are available. We will use the distribution
 function $P_N(\sigma)$ in the form proposed in \cite{BBFS,BBFS1}
\begin{equation}
P_N(\sigma)=N(a,n)\frac{\sigma/\sigma_0}{\sigma/\sigma_0+a}
e^{-(\sigma-\sigma_0)^{n}/(\Omega\sigma_0)^{n}}.
\end{equation}
This distribution function has been considered for the three cases:
 $n$=2, 6, 10 and fitted to the  characteristic value $\kappa_2$=0.25.
 The quantities $\langle\sigma^3\rangle/\langle\sigma\rangle^3$,
$\langle\sigma^4\rangle/\langle\sigma\rangle^4$, $\langle\sigma^5\rangle/
\langle\sigma\rangle^5$ which enter into (14) are significantly different
 from 1. For example for $n$=2:
$\langle\sigma^3\rangle/\langle\sigma\rangle^3$=1.82,
 $\langle\sigma^4\rangle/\langle\sigma\rangle^4$=2.97,
$\langle\sigma^5\rangle/\langle\sigma\rangle^5$=5.3.

 To reduce the influence of the specific type of the distribution function,
it is handy to introduce the relative cumulants \cite{BF}:

\begin{equation}
\kappa_3=\frac{\langle(\sigma-\langle\sigma\rangle)^3\rangle}
{\langle\sigma\rangle^3},\quad
\kappa_4=\frac{\langle(\sigma-\langle\sigma\rangle)^4\rangle}
{\langle\sigma\rangle^4},\quad
\kappa_5=\frac{\langle(\sigma-\langle\sigma\rangle)^5\rangle}
{\langle\sigma\rangle^5}.
\end{equation}
It allows us to rewrite the expression for $r$ in a form which
is less sensitive to the particular choice of the distribution
function $P_N(\sigma)$, since the relative cumulants are  small:
\begin{displaymath}
\begin{tabular}{|c|c|c|} \hline
\ $n$=2, $\Omega$=1.5, $a$=1& \ $n$=6, $\Omega$=1.1, $a$=0.1& \ $n$=10,
$\Omega$=11, $a$=1\\
\hline
$\kappa_3=0.07$ & $\kappa_3=0 $& $\kappa_3=0$\\
$\kappa_4=0.19$ & $\kappa_4=0.13$ & $\kappa_4=0.13$\\
$\kappa_5=0.15$ & $\kappa_5=0.02$&$\kappa_5=0$\\
\hline
\end{tabular}
\end{displaymath}
\\
Making use of the cumulants, $r$ can be presented as
\begin{eqnarray}
r&=&16-12\gamma+4.25\gamma^2-0.875\gamma^3+\kappa_2(-0.563\gamma^2+
0.375\gamma^3)\nonumber\\
&-&\frac{\kappa_3}{\kappa_2}(6\gamma-4.25\gamma^2+1.34\gamma^3-
0.125\kappa_2\gamma^3)+\frac{\kappa_4}{\kappa_2}(1.23\gamma^2-
0.75\gamma^3)-0.156\gamma^3\frac{\kappa_5}{\kappa_2}.
\end{eqnarray}

 At this point let us examine the experimental data. We use here
 information on proton diffractive dissociation from $^4$He \cite{Bujak}
  and proton \cite{Kuzn} targets
 at small momentum transfer and average
energy 300 GeV. These are the only experimental data on inclusive
coherent diffraction  of nuclei existing at the moment.
To the best of our knowledge they were overlooked for many years.
Using proposed  exponential parameterization for the cross-sections at small
 momentum transfer, we extrapolated data to $t$=0.
 In both cases the differential cross-section $d^2\sigma\Big{/}dtdM^2$ was
 integrated over the region $2.5(\rm{GeV})^2<\rm{M}^2<8(\rm{GeV})^2$ for
 which experimental data are available. The value of
the ratio of differential cross-sections, $r$,
extracted from experimental data, is found to be:

\begin{equation}
r=\int_{2.5({\rm GeV})^2}^{8({\rm GeV})^2}\,\Big(\frac{d\sigma^{pHe}}{dM^2dt}
\Big)_{t=0}dM^2\Bigg/\int_{2.5({\rm GeV})^2}^{8({\rm GeV})^2}\,
\Big(\frac{d\sigma^{pp}}{dM^2dt}\Big)_{t=0}dM^2=7.1\pm 0.7.
\end{equation}

The main error comes from the procedure of extrapolation to
$t$=0.

Note that the theoretical value of $r$ implies the integration over all
 diffractive masses~M$^2$. The available data cover most of the interval.
 The error coming from the fact that the integrations in eq. (18) is
 performed over the interval excluded the part of diffractive masses~M$^2$
 is small. Moreover, we checked that the ratio $\Big(\frac{d\sigma^{pHe}}
{dM^2dt}\Big)_{t=0}\Bigg/\Big(\frac{d\sigma^{pp}}{dM^2dt}\Big)_{t=0}$
depends on M$^2$ weakly. Hence the ratio of the small corrections to the
cross-sections $\Big(\frac{d\sigma^{pHe}}{dM^2dt}\Big)_{t=0}$ and
 $\Big(\frac{d\sigma^{pp}}{dM^2dt}\Big)_{t=0}$ originating from
 the region not covered experimentally is approximately equal to
$r$ which makes the error even smaller.  Thus the  correction related
to the inclusion of all diffractive  masses in the integrals in
eq. (18) is small as compared to the main error.

 The extracted value of $r$ should be compared with $r$=16 which one
would expect in the   impulse approximation.

 For the available region of M$^2$ the parameter $\beta$ describing the
 amplitude of diffractive scattering  was found to be 8$\pm$1 (GeV/c)$^{-2}$.
 We should use this value together with $\beta=13 \pm 0.5$
(GeV/c)$^{-2}$ corresponding to elastic proton-proton scattering because the
 process of interest includes both types of nucleon interaction. We notice
that the diagrams of the studied process contain an equal number of vertices
 of both types. Hence, it seems natural to use for $\beta$ the mean value of
 these two values.

 We found $\beta=10.5 \pm $0.6(\rm{GeV/c})$^{-2}$ and $\gamma=1.01\pm 0.02$.
With these results we can now  present $r$ as (we do not give here the errors
for the coefficients since the errors are correlated)
\begin{equation}
r=7.32-0.19\kappa_2-\frac{\kappa_3}{\kappa_2}(3.1-0.13\kappa_2)+
0.51\frac{\kappa_4}{\kappa_2}-0.16\frac{\kappa_5}{\kappa_2}.
\end{equation}
One can see from eq.(19) that for the small values of cumulants
the result is mainly sensitive to the value of $\kappa_3/\kappa_2$.
Thus  our theoretical predictions for $r$,

\begin{eqnarray}
\underline{n=2}\quad r&=&6.79\pm0.13\nonumber\\
\underline{n=6}\quad r&=&7.53\pm0.13\nonumber\\
\underline{n=10}\quad r&=&7.54\pm0.13,
\end{eqnarray}
differ mainly due to the
different values of $\kappa_3/\kappa_2$ for $n=2$ and for $n=6,10$.
The  range given by eq.(20), $r= 6.8 \div 7.6$,  is consistent
with $r=7.1\pm0.7$
extracted from the data.

\section{conclusion}

We  calculated
  the differential cross section~$d\sigma/dt$ of the coherent
diffractive dissociation of protons off $^4$He   at zero momentum transfer
and at high energy  in terms of the relative cumulants of the
distribution $P_N(\sigma)$. We found that the data are sensitive
to the moments $\int \sigma^n P_N(\sigma) d \sigma $ up to n=5.
 Current models
of $P_N(\sigma)$  describe the $^4$He data with an accuracy of
about 10\%.
The data support the small value of the ratio $\kappa_3/
\kappa_2$ already indicated by an analysis \cite{BBFS1} of the deuteron data
as well as a rather large dispersion of $P_N(\sigma)$
around the mean value $\sigma$.
Clearly new measurements of the $t$-dependence
 of the cross-section of diffractive dissociation at small $t$
using $^4$He and $^3$He are necessary using modern jet targets. They
would
 allow one to separate different terms in eq.(10) and   would
significantly improve sensitivity to details of the distribution $P_N(\sigma)$.

We thank L.Frankfurt and G.A.Miller for useful discussions. This work was
 partially supported by the U.S.DOE.
\bibliographystyle{unsrt}

\end{document}